\documentclass[%
 amsmath,amssymb,
 aps,
]{revtex4-2}

\usepackage[pdftex]{graphicx}
\usepackage{dcolumn}
\usepackage{bm}
\usepackage{comment}
 \usepackage{subfigure}
 \usepackage{here}

\begin{document}

\preprint{APS/123-QED}

\title{Bayesian Spectral Deconvolution of X-Ray Absorption Near Edge Structure Discriminating High- and Low-Energy Domains}

\author{Shuhei Kashiwamura$^1$, Shun Katakami$^2$, Ryo Yamagami$^3$, Kazunori Iwamitsu$^4$, \\Hiroyuki Kumazoe$^5$,  Kenji Nagata$^6$, Toshihiro Okajima$^7$, Ichiro Akai$^5$, Masato Okada$^{1,2}$}
 \email{okada@edu.k.u-tokyo.ac.jp}
\affiliation{%
$^1$Graduate School of Science, The University of Tokyo, Bunkyo, Tokyo 113-0033, Japan
$^2$Graduate School of Frontier Sciences, The University of Tokyo, Kashiwa, Chiba 277-8561, Japan\\
$^3$Graduate School of Science and Technology, Kumamoto University 860-8555, Japan\\
$^4$Technical Division, Kumamoto University, Kumamoto 860-8555, Japan\\
$^5$Institute of Industrial Nanomaterials, Kumamoto University, Kumamoto 860-8555, Japan\\
$^6$Research and Services Division of Materials Data and Integrated System, National Institute for Materials Science, Tsukuba, Ibaraki 305-0047, Japan\\
$^7$Aichi Synchrotron Radiation Center, Seto, Aichi 489-0965, Japan
}%





\begin{abstract}
In this paper, we propose a Bayesian spectral deconvolution considering the properties of peaks in different energy domains. Bayesian spectral deconvolution regresses spectral data into the sum of multiple basis functions. 
Conventional methods use a model that treats all peaks equally.
However, in X-ray absorption near edge structure (XANES) spectra, the properties of the peaks differ depending on the energy domain, and the specific energy domain of XANES is essential in condensed matter physics.
We propose a model that discriminates between the low- and high-energy domains.
We also propose a prior distribution that reflects the physical properties.
We compare the conventional and proposed models in terms of computational efficiency, estimation accuracy, and model evidence. We demonstrate that our method effectively estimates the number of transition components in the important energy domain, on which the material scientists focus for mapping the electronic transition analysis by first-principles simulation.
\end{abstract}

\newcommand{\argmax}{\mathop{\rm arg~max}\limits}
\newcommand{\argmin}{\mathop{\rm arg~min}\limits}
\maketitle

\section{Introduction}


X-ray absorption near edge structure (XANES) appears in the energy region of $\sim$50~eV around the X-ray absorption edge. It comes from the X-ray absorption due to electronic transitions from a core level to unoccupied band states for the element selected by the edge energy of X-ray absorption.\cite{nexafs}
XANES spectra are measured in condensed matter research because they offer crucial physicochemical information. They markedly change with the atom valence, the crystal field dependent on microscopic symmetry, and chemical bonding states. \cite{de2008core}
The experimental results of XANES are compared with the density of states (DoS)\cite{kotobuki, database} obtained by first-principles calculations, and the electronic states are discussed.

 
  XANES spectra contain multiple peaks and step structures. It is necessary to decompose the spectrum into these components to extract information about electronic transitions. However, estimating the number of peaks and each peak's width, position, and intensity is generally challenging since the spectrum is often complex and multimodal. A naive approach is least-squares fitting by the steepest descent method, but it cannot avoid the local minima.\cite{major2020practical} 
  Local minima show results that depend on the initial values, making the analysis arbitrary.
 
 
 An effective approach for the analysis of XANES spectra is Bayesian spectral deconvolution\cite{iwamitsu2020}, where the following spectral parameters can be estimated: position, width, and intensity of the basis function. \cite{nagata2012}
In  previous studies,\cite{nagata2012, iwamitsu2020} the exchange Monte Carlo method (EMC)\cite{hukushima1996} was used to get out of the local solution and approach the optimal global solution. EMC makes it possible to sample from the posterior distribution of spectral parameters.
The number of peaks and noise variance can also be estimated by maximizing a function called the marginal likelihood.\cite{tokuda2017}


 Previous studies on Bayesian spectral deconvolution used the regression function treating all peaks equally.\cite{nagata2012, tokuda2017, iwamitsu2020}
 However, the properties and physical background of XANES spectra differ in the high- and low-energy domains.
 Conventional methods have not fully utilized these differences.

In this work, we propose the energy-domain-aware regression model for Bayesian spectral deconvolution. It incorporates the prior knowledge that XANES spectra have sharp peaks in the low-energy domain and broad peaks in the high-energy domain.
 Our proposed model enables us to discuss the number of peaks in each energy domain separately and shows its superiority in convergence, sampling efficiency, estimation accuracy, and model evidence compared with the conventional model. This was confirmed via synthetic data analysis.


This paper is organized as follows.  
In Sect. 2, we introduce the framework of the Bayesian spectral deconvolution. We describe our model and show that our model contains the conventional model depending on the design of prior distributions. In Sect. 3, we compare the conventional and proposed models and verify our framework via synthetic data analysis. In Sect. 4, we present our discussion and conclude this paper. 
\section{Framework}
\begin{figure}[t]
 \begin{center}
  \subfigure[]{
   \includegraphics[width=75mm]{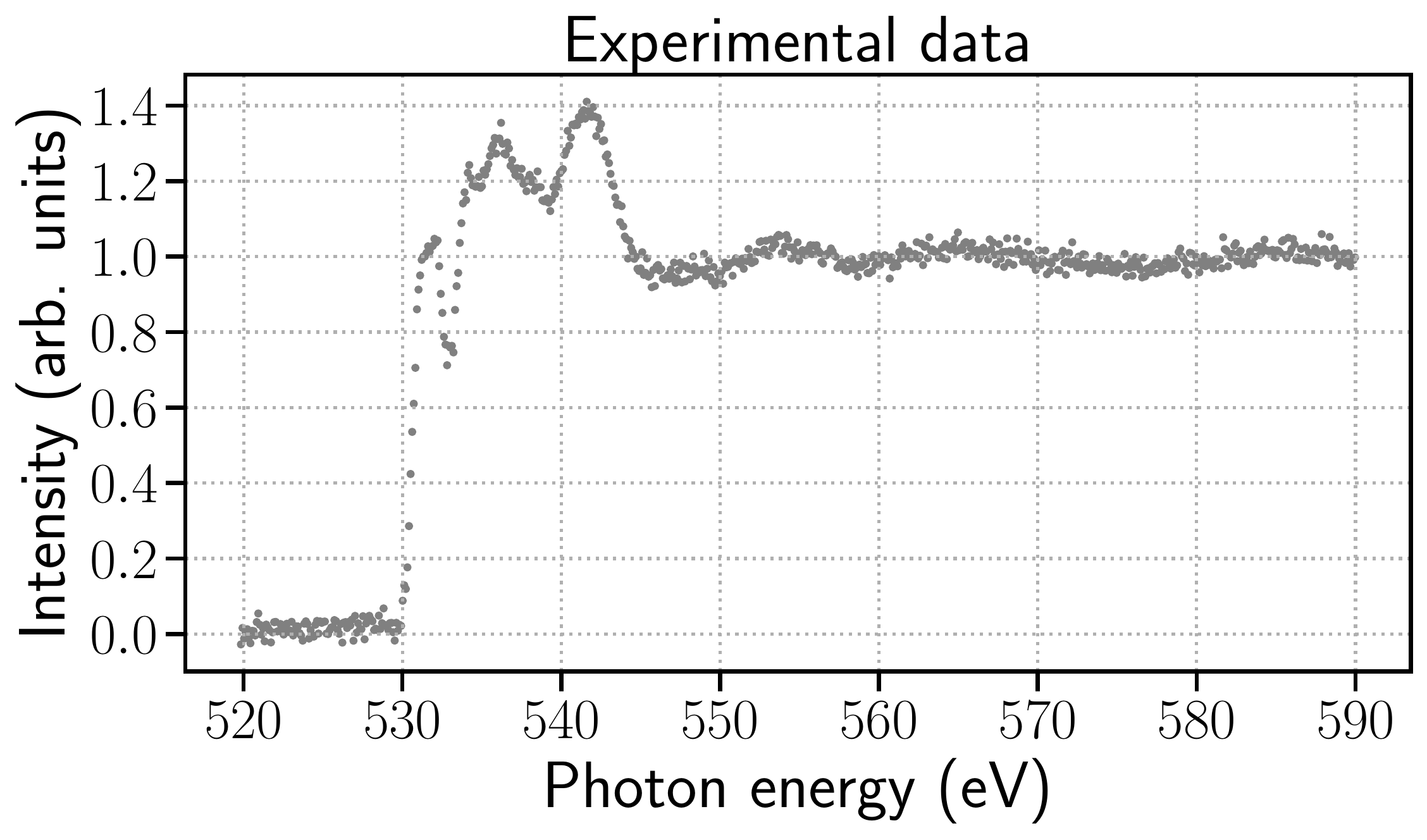}
   \label{1(a)}
  }~
  \subfigure[]{
   \includegraphics[width=75mm]{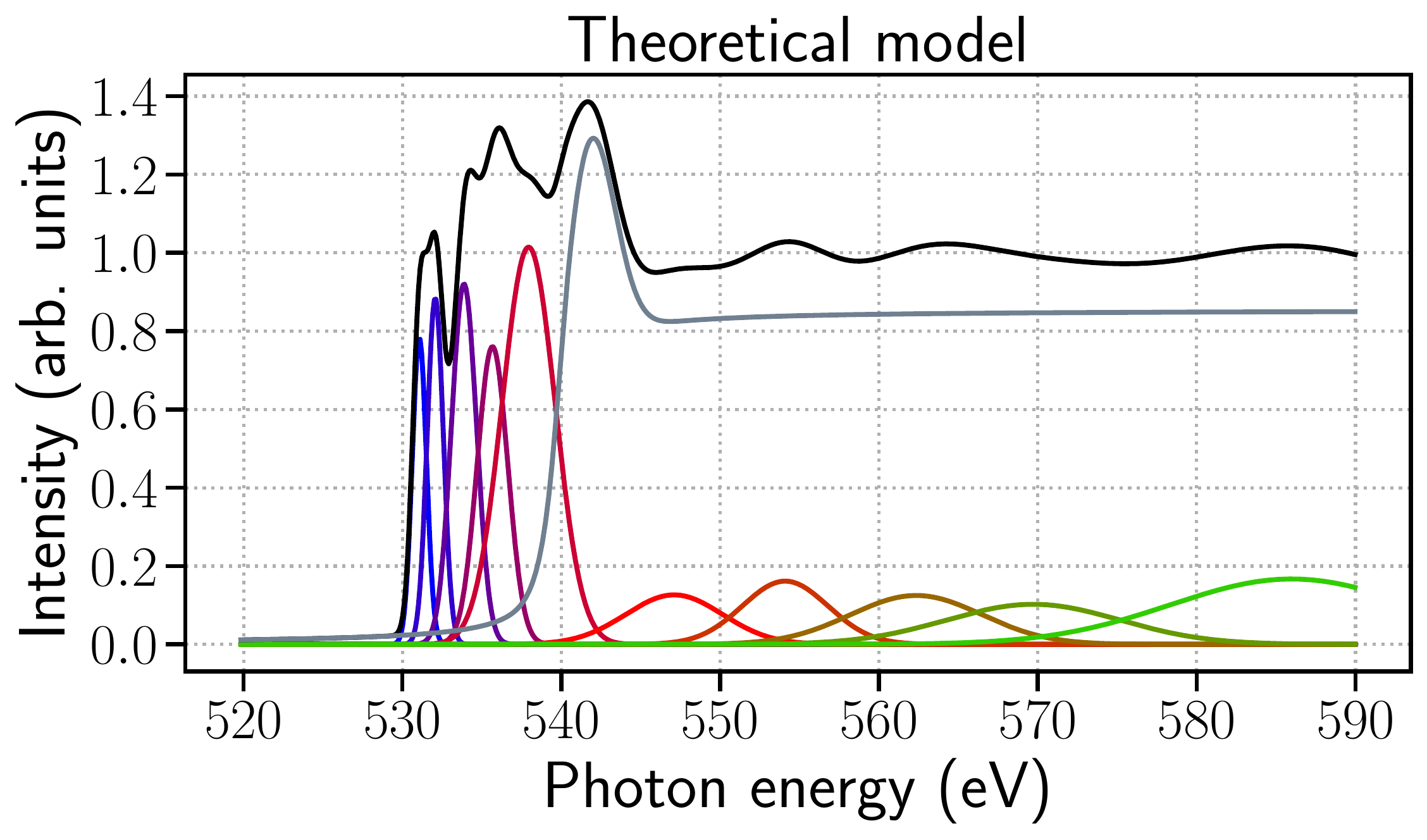}
   \label{1(b)}
  }
  \caption{
Synthetic model and data used in our analysis. (a) Gray dots show synthetic data. (b) Graph of spectrum model (black line) consisting of individual Gaussian peaks (colored lines) and the absorption edge with white line (gray line). }
  \label{syntheticdata}
 \end{center}
\end{figure}
\subsection{Model}

In XANES spectra, peaks in the low-energy domain reflect discrete electronic transitions from the atomic core levels to unoccupied states or valence bands.\cite{nexafs} In contrast, peaks in the high-energy domain reflect transitions to continuum states, also observed in extended X-ray absorption fine structure (EXAFS). \cite{exafs}
Thus, peaks in the low-energy domain are focused for mapping to the electronic transition analysis by first-principles simulation.
The shape of the peak reflects the transition probability and density of states. 
In the high-energy domain, the probability of electronic transitions is low because the nucleus weakly affects the continuous level.
Large sharp peaks tend to be seen in the low-energy domain, and small, broad peaks tend to be seen in the high-energy domain.\cite{nexafs} 


Figure \ref{syntheticdata} shows the synthetic data and model.
Figure \ref{1(a)} shows an example of the synthetic spectrum generated assuming XANES data measured at the O-K edge ($543.1\, \mathrm{eV}$) of $\rm{Li_5La_3Ta_2O_{12}}$, which has a complex structure in the low-energy domain.
$\rm{Li_5La_3Ta_2O_{12}}$ is expected to be used as a solid electrolyte material.\cite{randy}
Figure \ref{1(b)} shows the spectrum components; the XANES spectrum consists of peaks and a step structure called the absorption edge. At least one of the peaks appears near the absorption edge and is called the white line.


We propose a model that discriminates the peaks below and above the absorption edge.
We assume that there are $K_1$ peaks below the absorption edge and $K_2$ peaks above the absorption edge.
The spectral function $f(E;\theta, \mathcal{K})$ for the incident energy value of $E \in \mathbb{R}$ can be written as 
\begin{eqnarray}
f(E;\theta, \mathcal{K}) &:=& f_{step}(E;\theta_{step}) + f_{peak}(E;\theta_{peak}, \mathcal{K}),\label{model}
\end{eqnarray}
where the first term represents the absorption edge and white line. The second term represents the $K_1 + K_2$ peaks, where $\mathcal{K} := (K_1,  K_2)$ is the parameter of the number of peaks.
$\theta = \left\{\theta_{step},  \theta_{peak}\right\}$ is the spectral parameter.
We use the  arctangent function for the model of the step structure.
The shape of peaks depends on the instrumental resolution. 
If the instrumental resolution is smaller than the intrinsic lifetime broadenings, peaks have the Lorentzian lineshape.
On the other hand, if the instrumental resolution dominates, peaks have a Gaussian lineshape.

Our framework is envisioned to be applied most recently to XANES data measured at the O-K edge in Saga Light Source (SAGA-LS), and we use the Gaussian for the model of peaks because the instrumental resolution dominates.
The following discussion does not lose its generality when the Gaussian is replaced by the Lorentzian or pseudo-Voigt function. 


The function $f_{step}(E;\theta_{step})$ is defined as
\begin{eqnarray}
f_{step}(E;\theta_{step}) &:=& H\left\{\frac{1}{2} +  \frac{1}{\pi}\arctan\left(\frac{E - E_0}{\Gamma / 2}\right) \right\}
\nonumber \\
&&+ A \exp\left\{-4 \ln 2\left( \frac{E - (E_0 +\Delta E)}{\omega}\right)^2\right\}, \label{step_wl}
\end{eqnarray}
where $\theta_{step} =  \{H,E_0,\Gamma ,A,\Delta E,\omega\}$ is the parameter set.

Hartree atomic units are used here.
Here, $H$, $E_0$, and $\Gamma$ are the step's height, position, and width, respectively.
$A$, $E_0 + \Delta E$, and $\omega$ are the white line's height, position, and width, respectively.


The function of $K_1 + K_2$ peaks is defined as 
\begin{eqnarray}
f_{peak}(E;\theta_{peak}, \mathcal{K} = (K_1, K_2))&:=& f_1(E;\theta_1, K_1) + f_2(E;\theta_2, K_2) ,\label{f_peak}\\
f_1(E;\theta_1, K_1) &:=& \sum_{k_1=1}^{K_1} F_{k_1} \exp \left\{-4\ln2\left(\frac{E-E_{k_1}}{W_{k_1}}\right)^2\right\}, \\
f_2(E;\theta_2, K_2) &:=& \sum_{k_2=1}^{K_2} F_{k_2} \exp \left\{-4\ln2\left(\frac{E-E_{k_2}}{W_{k_2}}\right)^2\right\} ,
\end{eqnarray}
where $\theta_{peak} = \left\{\theta_1, \theta_2\right\}$, $\theta_1 = \{F_{k_1}, E_{k_1}, W_{k_1}\}_{{k_1}=1}^{K_1}$, and $\theta_2 = \{F_{k_2}, E_{k_2}, W_{k_2}\}_{{k_2}=1}^{K_2}$ are parameters and represent each peak's height, position, and width, respectively.  The function $f_1(E;\theta_1, K_1)$ represents the $K_1$ peaks below the absorption edge and the function $f_2(E;\theta_2, K_2) $ represents the $K_2$ peaks above the absorption edge. The peak positions must satisfy the conditions $E_{k_1} < E_0$, $E_{k_2} > E_0$.

We assume that an observed spectral intensity $I$ is obtained as the sum of model $f(E;\theta, \mathcal{K})$ and noise $\epsilon$ as
\begin{eqnarray}
I &=&  f(E;\theta, \mathcal{K})+ \epsilon. \label{sto}
\end{eqnarray}
We assume that noise $\epsilon$ is generated from a Gaussian distribution with zero mean and variance $1/b$. The conditional probability of observing $I$ given model $f(E; \theta, \mathcal{K})$ is then given by 

\begin{eqnarray}
p(I | E, \theta, \mathcal{K} ,b) &=& \left(\frac{b}{2 \pi} \right)^{\frac{1}{2}} \exp \left\{ -\frac{b}{2} \left[I - f(E ; \theta, \mathcal{K})\right]^2 \right\},
\end{eqnarray}
where $b$ is called precision.
Assuming that we have a data set $D := \{ E_i, I_i\}^N_{i=1}$ consisting of $N$ observed data, the conditional probability $p(D|\theta, \mathcal{K}, b)$ is given by 
\begin{eqnarray}
p(D|\theta, \mathcal{K}, b) &=& \prod _{i=1}^N p(I_i|E_i,\theta, \mathcal{K}, b) \\
&=& \left(\frac{b}{2 \pi}\right)^{\frac{N}{2}} \exp (-Nb \mathcal{E}_N(\theta, \mathcal{K})),
 \end{eqnarray}
where $\mathcal{E}_N(\theta, \mathcal{K})$ is the error function of the fitting function $f(E ; \theta, \mathcal{K})$:
\begin{eqnarray}
\mathcal{E}_N(\theta, \mathcal{K}) &=& \frac{1}{2N} \sum_{i=1}^N \left[I_i - f(E_i;\theta, \mathcal{K})\right]^2.
\end{eqnarray}

\subsection{Bayesian formulation}
Bayesian spectral deconvolution treats the data set $D$, spectral parameter $\theta$, the parameter of the number of peaks $\mathcal{K}$, and the inverse noise variance $b$ as random variables.
We assume that $\mathcal{K}$ and $b$ are generated subject to the probabilities $p(\mathcal{K})$ and $p(b)$, respectively, $\theta$ is generated subject to the probability $p(\theta | \mathcal{K})$, and data set $D$ is generated subject to the probability $p(D|\theta, \mathcal{K},b)$. The joint probability density $p(D,\theta, \mathcal{K}, b)$ is then given by

\begin{eqnarray}
p(D,\theta, \mathcal{K}, b) = p(D|\theta, \mathcal{K}, b)p(\theta|\mathcal{K})p(\mathcal{K})p(b). \label{joint}
\end{eqnarray}
The posterior distribution of data set $\theta$ given $D, \mathcal{K}$ and $b$ is represented by Bayes' theorem as
\begin{eqnarray}
p(\theta|D, \mathcal{K},b) &=&\frac{p(D,\theta, \mathcal{K}, b) }{\int d \theta p(D,\theta, \mathcal{K}, b) }\\
&=&\frac{p(D|\theta, \mathcal{K}, b)p(\theta|\mathcal{K})}{\int d \theta p(D|\theta, \mathcal{K}, b)p(\theta|\mathcal{K})} \\
 &=& \frac{1}{Z_N(\mathcal{K}, b)} \left(\frac{b}{2 \pi}\right)^{\frac{N}{2}} \exp\left[-Nb\mathcal{E}_N(\theta)\right] p(\theta |\mathcal{K}), \label{posteri}\\
Z_N(\mathcal{K},b) &:=& \int d \theta \left(\frac{b}{2 \pi}\right)^{\frac{N}{2}} \exp\left[-Nb\mathcal{E}_N(\theta)\right] p(\theta |\mathcal{K}), \label{marginalizedL}
\end{eqnarray}
where $Z_N(\mathcal{K},b)$ is marginal likelihood, which is also called model evidence.
The negative logarithm of $Z_N(\mathcal{K}, b)$ is called as the Bayesian free energy:
\begin{eqnarray}
F_N(\mathcal{K}, b) := -\log Z_N(\mathcal{K}, b)\label{free_energy}.
\end{eqnarray}


\subsection{Prior distributions}
The probability density of the spectral parameter $\theta$ given the parameter $\mathcal{K}$ is called the prior distribution.
We show the discrimination of the peaks below and above the absorption edge using a prior distribution.
We also show that, depending on the prior distribution, our model includes a conventional model that treats all peaks equally.\cite{nagata2012, tokuda2017, iwamitsu2020}

In the following, we define the probability density functions of prior distributions of each parameter. 
In this paper, $\text{U}(X;\alpha, \beta)$ denotes the uniform distribution defined in $\alpha \leq x \leq \beta$, 
$\text N(X;\mu, \sigma^2)$ denotes the normal distribution with the mean $\mu$ and the covariance $\sigma$, and $\text{G} (X;\kappa, \vartheta)$ denotes the gamma distribution with the shape parameter $\kappa$ and the scale parameter $\vartheta$ as  
\begin{eqnarray}
\text{U}(X;\alpha, \beta) &=& \frac{1}{\beta - \alpha} \,\,\,\,\,\,\,(\alpha < X <\beta), \\
\text{N} (X; \mu, \sigma^2) &=& \frac{1}{\sqrt{2 \pi \sigma^2}} \exp \left\{-\frac{1}{2\sigma^2}(X-\mu)^2\right\}, \\
\text{G} (X;\kappa, \vartheta) &=& \frac{1}{\Gamma(\kappa) \vartheta^{\kappa}} X^{\kappa-1} e^{-X/\vartheta},
\end{eqnarray}
where we refer  to $\alpha, \beta, \mu, \sigma, \kappa$, and $\vartheta$ as hyperparameters. 
\subsubsection{Proposed model}
We define the relationship between the position of the absorption edge $E_0 \in \theta_{step}$ and those of peaks $E_{k_1}, E_{k_2} \in \theta_{peak}$ as 
\begin{eqnarray}
\Delta E_{k_1} &:=& E_{k_1} -E_0 , \label{DeltaE_1}\\
\Delta E_{k_2} &:=& E_{k_2} -E_0, \label{DeltaE_2}
\end{eqnarray}
where we can rewrite the conditions $E_{k_1} < E_0$ and $E_{k_2} > E_0$ as $\Delta E_{k_1} <0$ and $\Delta E_{k_2} > 0$, respectively.
The parameters $\theta_{step}$ and $\theta_{peak}$ are not independent variables because $E_{k_1} \in \theta_{peak}$ and $E_{k_2} \in \theta_{peak}$ depend on $E_0 \in \theta_{step}$ via Eqs. (\ref{DeltaE_1}) and (\ref{DeltaE_2}).
To treat parameters independently, we reparametrize $\theta_{peak}, \theta_1,$ and $\theta_2$ as $\theta'_{peak} := \left\{\theta'_1, \theta'_2\right\}$,
$\theta'_1 := \{F_{k_1}, \Delta E_{k_1}, W_{k_1}\}_{{k_1}=1}^{K_1}$, and $\theta'_2 := \{F_{k_2}, \Delta E_{k_2}, W_{k_2}\}_{{k_2}=1}^{K_2}$.
Then, the prior distribution $p(\theta | \mathcal{K})$ can be rewritten as 
\begin{eqnarray}
p(\theta | \mathcal{K}) &&= p(\theta_{step}, \theta_{peak} | \mathcal{K}) , \label{joint}\\
&& = p(\theta_{peak} | \theta_{step}, \mathcal{K})p(\theta_{step} | \mathcal{K}) ,\\
&& = p(\theta'_{peak} | \mathcal{K})p(\theta_{step}),
\end{eqnarray}
where $\theta_{step}$ and $\theta'_{peak}$ are independent variables.


\begin{table}[th]
\centering
\caption{Prior distributions for absorption edge and white line.}
\begin{tabular}{cccc}\hline
spectral parameter & prior & hyperparameter & \\ \hline
$H$                  & $p(H):=\text{U}(H;\alpha_H, \beta_H)$&$\alpha_H ,\,\, \beta_H $          \\
$E_0$               &$p(E_0):=\text{N} (E_0; \mu_{E_0}, \sigma_{E_0}^2)$ &$\mu_{E_0} ,  \,\, \sigma_{E_0} $           \\
$\Gamma$             & $p(\Gamma):=\text{U}(\Gamma;\alpha_{\Gamma}, \beta_{\Gamma})$&$\alpha_{\Gamma} , \,\, \beta_{\Gamma} $              \\
$A$                  &  $p(A):=\text{G} (A;\kappa_A, \vartheta_A)$     &$\kappa_A $, \,\, $\vartheta_A $        \\
$\Delta E$           & $p(\Delta E):=\text{N} (\Delta E; \mu_{\Delta E}, \sigma_{\Delta E}^2)$ &$\mu_{\Delta E}=0 , \,\, \sigma_{\Delta E}$      \\
$\omega $             &$p(\omega):=\text{U}(\omega;\alpha_{\omega}, \beta_{\omega})$ &$\alpha_{\omega} , \,\, \beta_{\omega} $    \\ \hline         
\end{tabular}
\label{table_one}
\end{table}

\begin{table}[th]
\centering
\caption{Prior distributions for peaks in proposed model.}
\begin{tabular}{cccc}\hline
spectral parameter & prior & hyperparameter & \\ \hline
$F_{k_1}$                  & $p(F_{k_1}):=\text{G}(F_{k_1};\kappa_{F_{k_1}}, \vartheta_{F_{k_1}})$&$\kappa_{F_{k_1}} ,\,\, \vartheta_{F_{k_1}} $          \\
$F_{k_2}$               &$p(F_{k_2}):=\text{G}(F_{k_2};\kappa_{F_{k_2}}, \vartheta_{F_{k_2}})$&$\kappa_{F_{k_2}} ,\,\, \vartheta_{F_{k_2}} $          \\
$\Delta E_{k_1}$             & $p(\Delta E_{k_1}):=\text{U}(\Delta E_{k_1};\alpha_{\Delta E_{k_1}}, \beta_{\Delta E_{k_1}})$&$\alpha_{\Delta E_{k_1}} , \,\, \beta_{\Delta E_{k_1}} $              \\
$\Delta E_{k_2}$                  &  $p(\Delta E_{k_2}):=\text{U}(\Delta E_{k_2};\alpha_{\Delta E_{k_2}}, \beta_{\Delta E_{k_2}})$&$\alpha_{\Delta E_{k_2}} , \,\, \beta_{\Delta E_{k_2}} $              \\
$W_{k_1}$           & $p(W_{k_1}):=\text{G}(W_{k_1};\kappa_{W_{k_1}}, \vartheta_{W_{k_1}})$&$\kappa_{W_{k_1}} ,\,\, \vartheta_{W_{k_1}} $          \\
$W_{k_2} $             &$p(W_{k_2}):=\text{G}(W_{k_2};\kappa_{W_{k_2}}, \vartheta_{W_{k_2}})$&$\kappa_{W_{k_2}} ,\,\, \vartheta_{W_{k_2}} $          \\ \hline         
\end{tabular}
\label{table_two}
\end{table}
First, we define the prior distributions on the absorption edge and the white line.
 The prior distribution of $\theta_{step}$ can be written as the product of the prior distribution of each parameter: 
\begin{eqnarray}
p(\theta_{step}) &=& p(H)p(E_0)p(\Gamma)p(A)p(\Delta E)p(\omega) .
\end{eqnarray}
We show definitions of the prior distributions $p(H), p(E_0), p(\Gamma), p(A), p(\Delta E)$, and $p(\omega)$ in Table \ref{table_one}.
Since the position of the absorption edge $E_0$ appears near the ionization energy $\mu_{E_0}$ of the element of interest, we set up a Gaussian distribution centered on $\mu_{E_0}$ as $p(E_0)$.
Since the white line position $E_0 + \Delta E$ is near the absorption edge, we use a Gaussian distribution centered on $\mu_{\Delta E} := 0$ as $p(\Delta E)$.
We use the gamma distribution for the intensity parameter of white line $A$ to prevent the intensity becoming too small.
We set the uniform distribution for $H, \Gamma$, and $\omega$ since we have no knowledge except for a rough range.

Next, we define the prior distributions of peaks.
We can write the prior distribution of $\theta'_{peak}$ as 
\begin{eqnarray}
p(\theta'_{peak}) &=& p(\theta'_1 | K_1)p(\theta'_2 | K_2), \\
p (\theta'_1|K_1) &=& \prod_{k_1=1}^{K_1} p(F_{k_1})p(\Delta E_{k_1})p(W_{k_1}), \\
p (\theta'_2|K_2) &=& \prod_{k_2=1}^{K_2} p(F_{k_2})p(\Delta E_{k_2})p(W_{k_2}).
\end{eqnarray}
We show definitions of the prior distributions $p(F_{k_1}), p(F_{k_2}), p(\Delta E_{k_1}), p(\Delta E_{k_2}), p(W_{k_1})$, and $p(W_{k_2})$ in Table \ref{table_two}.
We restrict the positions of peaks by the uniform distribution.
Assuming that the range of the peak positions is $E_{min} \leq E \leq E_{max}$, we set hyperparameters $\alpha_{\Delta E_{k_1}}, \beta_{\Delta E_{k_1}}, \alpha_{\Delta E_{k_2}}$, and $\beta_{\Delta E_{k_2}}$ as
\begin{equation}
\begin{cases}
\; \alpha_{\Delta E_{k_1}} = E_{min}-E_0-\sigma_{E_0} \\
\;  \beta_{\Delta E_{k_1}} = 0
 \end{cases}  
 \begin{cases}
\; \alpha_{\Delta E_{k_2}} = 0 \\
\;  \beta_{\Delta E_{k_2}} = E_{max} - E_0 + \sigma_{E_0},
 \end{cases}  
\end{equation}
where we consider that $p(E_0)$ is the normal distribution with width $\sigma_{E_0}$.
These prior distributions satisfy the condition $\Delta E_{k_1} <0$, $\Delta E_{k_2} > 0$.
We use the gamma distribution for the prior distribution of the intensity and width.
Using the gamma distribution, we can restrict the parameters to non zero and control the shape of the distribution by adjusting the hyperparameters.
We can incorporate the prior distributions on the knowledge that $K_1$ peaks below $E_0$ are sharp and $K_2$ peaks above $E_0$ are broad.
The examples of concrete hyperparameters are shown in Sect. 3.
\begin{figure}[t]
 \centering
 \includegraphics[width=150mm]{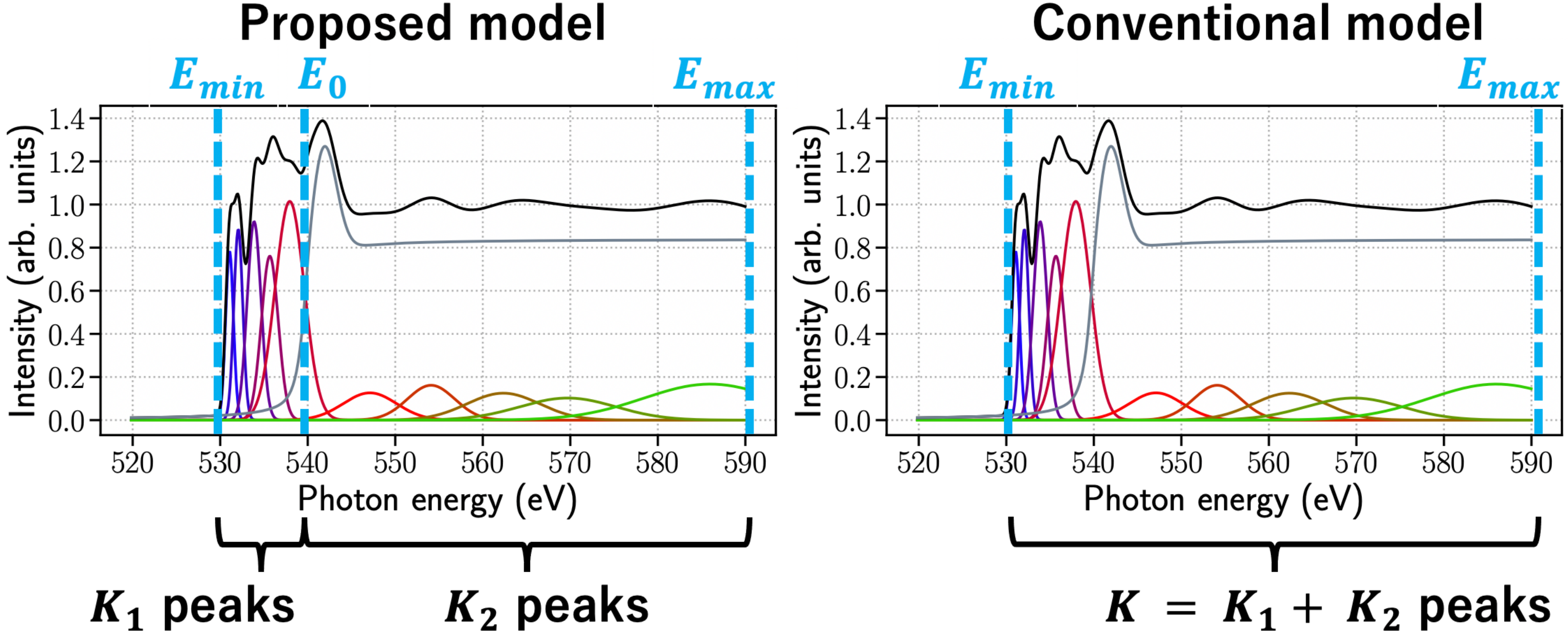}
 \caption{Illustration of conventional and proposed models.}
 \label{conventional_proposed}
\end{figure}

\subsubsection{Conventional model}
We show that our model can represent the conventional model by unifying the prior distributions below and above the absorption edge as
\begin{eqnarray}
p(F_{k_1}) &=& p(F_{k_2}), \\
p(E_{k_1}) &=& p(E_{k_2}), \\
p(W_{k_1}) &=& p(W_{k_2}).
\end{eqnarray}
Since the conventional model does not discriminate the peaks below and above the step, $E_0$ is independent of $E_{k_1}$ and $E_{k_2}$.
Thus, $\theta_{step}$ and $\theta_{peak}$ are independent and the right side of Eq. (\ref{joint}) can be written as
\begin{eqnarray}
p(\theta_{step}, \theta_{peak} | \mathcal{K}) = p(\theta_{step})p(\theta_{peak} | \mathcal{K}).
\end{eqnarray}
Here, all $K_1 + K_2$ peaks have the same conditions.
Thus, we do not need to set $K_1$ or $K_2$, and only determine $K := K_1 + K_2$, which corresponds to their sum.
We define the parameter of the number of peaks as $\mathcal{K} = K$ for the conventional model.

\begin{table}[t]
\centering
\caption{Prior distributions for peaks in conventional model.}
\begin{tabular}{cccc}\hline
spectral parameter & prior & hyperparameter & \\ \hline
$F_{k}$                  & $p(F_k):=\text{U}(F_{k};\alpha_{F_{k}}, \beta_{F_{k}})$&$\alpha_{F_{k}} , \,\, \beta_{F_{k}} $              \\
$E_{k}$               &$p(E_k):=\text{U}(E_{k};\alpha_{E_{k}}, \beta_{E_{k}})$&$\alpha_{E_{k}} , \,\, \beta_{E_{k}} $              \\
$W_{k}$             & $p(W_k):=\text{U}(W_{k};\alpha_{W_{k}}, \beta_{W_{k}})$&$\alpha_{W_{k}} , \,\, \beta_{W_{k}} $              \\ \hline    
\end{tabular}
\label{table_three}
\end{table}

Figure \ref{conventional_proposed} shows the relationship between the proposed and conventional models.
As seen in the figure, there are $K_1$ peaks in $E_{min} < E < E_0$ and $K_2$ peaks in $E_0 < E < E_{max}$ in the proposed model and $K = K_1 + K_2$ peaks in $E_{min} < E  <E_{max}$ in the conventional model.
We redefine $f_{peak}(E;\theta_{peak}, \mathcal{K})$ for the conventional model as 
\begin{eqnarray}
f_{peak}(E ; \theta_{peak}, \mathcal{K} = K) := \sum_{k=1}^K F_{k} \exp \left\{-4\ln2\left(\frac{E-E_{k}}{W_{k}}\right)^2\right\}.
\end{eqnarray}

The prior distributions of the peaks in the conventional model can be written as $p(\theta_{peak} | \mathcal{K}) = \prod_{k}^{K=K_1+K_2} p(F_k)p(E_k)p(W_k)$.
We show definitions of the prior distributions $p(F_k), p(E_k),$ and $p(W_k)$ in Table \ref{table_three}. 
As shown in Table \ref{table_three}, we use uniform distributions for all parameters.
We set hyperparameters for the prior distributions of the position of peaks $E_k$, as $\alpha_{E_k} = E_{min}$ and $\beta_{E_k} = E_{max}$.

Since XANES contains peaks with a wide variety of intensities and widths, the conventional model has no choice but to use a uniform distribution.
The proposed model distinguishes the peaks below and above the absorption edge by devising a prior distribution for the energy position. This enabled us to reflect the physical properties in accordance with the energy domain in the prior distribution design.

\subsection{Exchange Monte Carlo method}
The calculation of the posterior probability Eq. (\ref{posteri}) is generally difficult since it requires high dimensional integration as Eq. (\ref{marginalizedL}).
However, we can obtain an empirical posterior distribution from samples by EMC without calculating $Z_N(\mathcal{K},b)$. 
 
 We prepare the different $L$ inverse variances $\{b_l\}_{l=1}^L$ and posterior distributions $\{P(\theta_l | D, \mathcal{K}, b_l)\}_{l=1}^L$. We can obtain samples from the following joint density by EMC:
  \begin{eqnarray}
p(\theta_1, \theta_2, ...\theta_L|D, \mathcal{K}, b_1, b_2, ...b_L) &=& \prod_{l=1}^L p(\theta_l |D, \mathcal{K}, b_l ),
\end{eqnarray}
where the inverse variance satisfy the condition $0 = b_1 < b_2 < ...b_L $ and are called the inverse temperatures.
 In addition to Metropolis-Hasting sweeps at each $L$ Markov chain called replicas, EMC uses exchange moves between adjacent pairs of replicas.
 To satisfy a detailed balance, the swap is accepted with probability
\begin{eqnarray}
u(\theta_{l+1}, \theta_{l}, b_{l+1},  b_l) & :=& \min\left[1, \frac{p(\theta_{l+1}|D,\mathcal{K},b_l) p(\theta_l | D,\mathcal{K},b_{l+1})}{p(\theta_{l}|D,\mathcal{K},b_l) p(\theta_{l+1} | D,\mathcal{K},b_{l+1})}\right] \\
&=& \min \left[1, \exp\{N(b_{l+1} -b_l)(\mathcal{E}_N(\theta_{l+1}) -\mathcal{E}_N(\theta_l)) \}\right],
\end{eqnarray}
where the $\theta_l$ is the current state of each replica. 

At low temperatures, the posterior density is concentrated around the parameter $\theta$ that minimizes the error function $\mathcal{E}_N(\theta)$ and has complex local minima in the parameter space. 
As the temperature increases, the posterior distribution asymptotes to the prior distributions, enabling the Markov chain to explore the parameter space without becoming trapped in the local regions.
If the replicas at low temperatures are trapped in the local minima, the exchange move enables it to reach the global minima through exploration at higher temperatures.


Moreover, we can calculate the model evidence or the free energy from samples obtained by EMC.
Here, we define the auxiliary function: \cite{ogata1990, nagata2012,tokuda2017}
\begin{eqnarray}
\tilde{Z}_N (\mathcal{K}, b) &:=&  \int d \theta  \exp\left[-Nb\mathcal{E}_N(\theta)\right] p(\theta |\mathcal{K}).
\end{eqnarray}
Equations.(\ref{posteri}) and (\ref{marginalizedL}) can be rewritten as
\begin{eqnarray}
p(\theta | D, \mathcal{K}, b) &=& \frac{1}{\tilde{Z}_N(\mathcal{K}, b)} \exp\left[-Nb\mathcal{E}_N(\theta)\right] p(\theta |\mathcal{K}), \\
Z_N(\mathcal{K}, b) &=& \left(\frac{b}{2 \pi} \right)^{\frac{b}{2}} \tilde{Z}_N(\mathcal{K},b).
\end{eqnarray}
We can calculate $\tilde{Z}_N(\mathcal{K},b_l)$ as 
\begin{eqnarray}
\tilde{Z}_N(\mathcal{K},b_l) &=& \frac{\tilde{Z}_N(\mathcal{K}, b_{l})}{\tilde{Z}_N(\mathcal{K}, b_{l-1})} \times \cdots \times \frac{\tilde{Z}_N(\mathcal{K}, b_{2})}{\tilde{Z}_N(\mathcal{K}, b_{1})} \\
&=& \prod_{l'=1}^{l-1} \frac{\tilde{Z}_N(\mathcal{K}, b_{l' +1})}{\tilde{Z}_N(\mathcal{K}, b_{l'})}\\
&=& \prod_{l'=1}^{l-1}\left<\exp\left[-N(b_{l'+1}-b_{l'})\mathcal{E}_N(\theta_l')\right]\right>_{b_l'},
\end{eqnarray}
where $<...>_{b_l'}$ denotes the expectation over the probability distribution $p(\theta_l | D, \mathcal{K}, b_l)$.
This is the extended method of importance sampling. \cite{prml}
We can then obtain Bayes free energy as 
\begin{eqnarray}
F_N(\mathcal{K}, b_l) = -\frac{b_l}{2} \log\left(\frac{b_l}{2 \pi}\right) - \log \tilde{Z}_N(\mathcal{K},b_l).
\end{eqnarray}

\subsection{Parameter estimation}
Here, we describe the method to estimate the number of peaks, noise variance, and spectral parameters.
The estimation of $\mathcal{K}$ is called model selection, where the most probable model is chosen.\cite{prml}
The optimal values of $\mathcal{K}'$ and $b_{l'}$ can be obtained by the empirical Bayes approach as 
\begin{eqnarray}
(\mathcal{K}', l') &:=& \argmax_{\mathcal{K}, l} Z_N(\mathcal{K},b_l) \\
&=& \argmin_{\mathcal
{K},l} F_N(\mathcal{K}, b_l), \label{noise} \label{min_F}
 \end{eqnarray} 
 where we select the noise value from the discrete inverse temperature $\{b_l\}_{l=1}^L$.
 We will only mention that we can interpolate the free energy by the multihistogram method considering the continuity of noise variance.\cite{tokuda2017}
 
 Let $\{\theta_{l'}^{(m)}\}_{m=1}^M$ be EMC samples for the $l$-th replica, where $M$ is the number of EMC samples.
 The optimal value of spectrum parameter $\theta_{l'}^{(m')}$ can be obtained by the maximization of the posterior probability as
 \begin{eqnarray}
m' &=&\argmax_{m} p(\theta_{l'}^{(m)} | D, \mathcal{K}', b_{l'}), \\
&=& \argmax_{m} \left[ \exp\left[-Nb_{l'} \mathcal{E}_N(\theta_{l'}^{(m)})\right] p(\theta_{l'}^{(m)} |\mathcal{K})\right].
\end{eqnarray}
This technique is called maximum a posteriori (MAP) estimation.
From the aforementioned steps, we can estimate the number of peaks $\mathcal{K}'$, inverse noise variance $b_{l'}$, and spectrum parameter 
$\theta_{l'}^{(m')}$.
\begin{table}[H]
\centering
\caption{Hyperparameters of priors for absorption edge and the white line.}
\begin{tabular}{ccc}
\hline
spectral parameter & hyperparameter \\ \hline
$H$                  & $\alpha_H = 0.8,\,\, \beta_H = 0.9$          \\
$E_0$               & $\mu_{E_0} = 543.1,  \,\, \sigma_{E_0} = 2.0$           \\
$\Gamma$             & $\alpha_{\Gamma} =0.5, \,\, \beta_{\Gamma} = 1.4$              \\
$A$                  &       $\kappa_A = 2.6$, \,\, $\vartheta_A = 0.6$        \\
$\Delta E$           &  $\mu_{\Delta E} = 0.0, \,\, \sigma_{\Delta E} = 2.0$      \\
$\omega $             & $\alpha_{\omega} = 2.0, \,\, \beta_{\omega} = 4.0$    \\ \hline         
\end{tabular}
\label{table_four}
\end{table}

\begin{table}[H]
\centering
\caption{Hyperparameters of priors for peaks in the conventional model.}
\begin{tabular}{ccc}
\hline
spectral parameter & hyperparameter \\ \hline
$F_k$                  & $\alpha_{F_{k}} = 0.0, \,\, \beta_{F_{k}} = 1.4$         \\
$E_k$               & $\alpha_{E_{k}} = 530, \,\, \beta_{E_{k}} = 590$         \\
$W_k$             & $\alpha_{W_{k}} = 0.5, \,\, \beta_{W_{k}} = 15.0$   \\ \hline
\end{tabular}
\label{table_five}
\end{table}

\begin{table}[H]
\centering
\caption{Hyperparameters of priors for peaks in the proposed model.}
\begin{tabular}{ccc}
\hline
spectral parameter & hyperparameter \\ \hline
$F_{k_1}$                  & $\kappa_{F_{k_1}} = 2.6, \,\, \vartheta_{F_{k_1}} = 0.6$          \\
$F_{k_2}$               & $\kappa_{F_{k_2}} = 4.0, \,\, \vartheta_{F_{k_2}} = 0.1$          \\
$\Delta E_{k_1}$             & $\alpha_{E_{k_1}} = -15.1, \,\, \beta_{E_{k_1}} = 0.0$              \\
$\Delta E_{k_2}$                  &     $\alpha_{E_{k_2}} = 0.0, \,\, \beta_{E_{k_2}} = 48.9$       \\
$W_{k_2}$          &  $\kappa_{W_{k_1}} = 3.0, \,\, \vartheta_{W_{k_1}} = 1.0$     \\
$W_{k_2}$        & $\kappa_{W_{k_2}} = 11.0, \,\, \vartheta_{W_{k_2}} = 0.8$    \\ \hline         
\end{tabular}
\label{table_six}
\end{table}

\subsection{Probability of the number of peaks}
As previously described, we estimate the number of peaks and noise variance by the empirical Bayes approach. 
In addition, we can calculate the probability density of the number of peaks and noise variance by the hierarchical Bayes approach as 
\begin{eqnarray}
p(\mathcal{K},b|D) &=& \frac{\int d \theta  \, p(D, \theta, \mathcal{K}, b)}{\sum_{\mathcal{K}} \int db \int  d \theta  \, p(D, \theta, \mathcal{K}, b)} \\
 &=&\frac{\exp \left[-F_N(\mathcal{K}, b) \right]}{\sum_{\mathcal{K}} \int db\exp \left[-F_N(\mathcal{K}, b) \right]},\label{hiera_bayes}\\
p(\mathcal{K}|D) &=&\int db p(\mathcal{K}, b), \label{prob_peak}
\end{eqnarray}
where we assume that $p(\mathcal{K})$ and $p(b)$ are uniform distributions.
Note that it is also possible to restrict $p(\mathcal{K})$ or $p(b)$ on the basis of the results of first-principles calculations or the conditions of the experiment.
 As seen in Eq. (\ref{hiera_bayes}), the maximum point of $p(\mathcal{K}, b | D)$ is the equivalent to the minimum point of the free energy $F_N(\mathcal{K}, b)$.
We can obtain $p(K|D)$ and $p(K_1, K_2 | D)$ via Eq. (\ref{prob_peak}) since the parameter $\mathcal{K}$ corresponds to $(K_1, K_2)$ in the proposed model and $K$ in the conventional model.
Moreover, we can calculate $p(K_1 | D)$ and $p(K_2|D)$ for the proposed model by the principle of marginalization:
\begin{eqnarray}
p(K_1|D) &=& \sum_{K_2} p(K_1, K_2|D), \\
p(K_2|D) &=& \sum_{K_1} p(K_1, K_2|D).
\end{eqnarray}

Thus, we can calculate the probability densities of the number of peaks below and above the absorption edge separately. As previously mentioned, the $K_1$ peaks are more important than the $K_2$ peaks from the viewpoint of physical properties. Focusing on the crucial domain and discussing the number of peaks are impossible using the conventional model, but the proposed model makes it possible.

\section{Numerical Experiments}\label{Sect3}
In this section, we compare the conventional model and proposed models by numerical experiments on the synthetic data in Fig. \ref{syntheticdata}. 
The synthetic data was generated in accordance with Eq. (\ref{sto}) with the number of peaks $K= (K_1+K_2) = 10$, inverse variance of noise $b = 3000$, and $N=703$. 
In the synthetic data, $K_1 = 5$ peaks are located below the absorption edge, whereas $K_2=5$ peaks are located above. 
Since the absorption edge appears near the ionization energy, $E_0$ is chosen to be about $543.1$ eV \cite{543.1}.

For the EMC method, the number $L$ of replicas was set using a geometric progression as\cite{nagata2008}
 \begin{equation}
b_l=\left\{
  \begin{aligned}
  &\,3000 \cdot \xi ^{(l - L + 2)} &(l \neq 1)\\
  &\,0   &(l=1),
  \end{aligned}
\right.
\end{equation}
where we set $b_{L-2} = 3000$, which is the true value of the noise variance. 
The hyperparameters of the prior distributions are shown in Tables \ref{table_four}-\ref{table_six}.
Here, we define $50$ iterations of the metropolis update as one Monte Carlo step (MCS). 
We performed 60,000 MCSs and discarded the first 30,000 of them as the burn-in period.


 \begin{figure}[t]
 \begin{center}
  \subfigure[]{
   \includegraphics[width=.47\columnwidth]{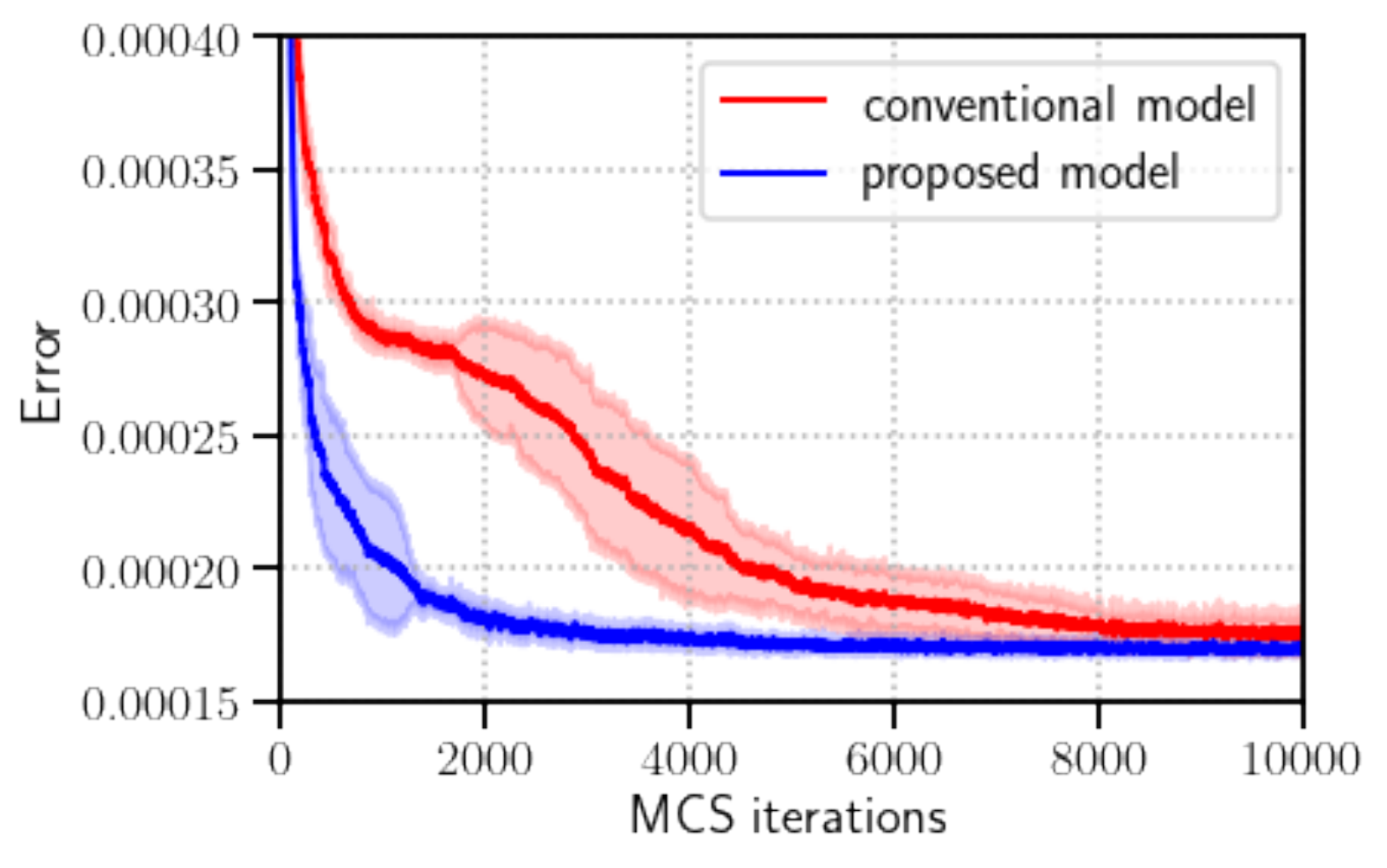}
   \label{3_a}
  }~
  \subfigure[]{
   \includegraphics[width=.47\columnwidth]{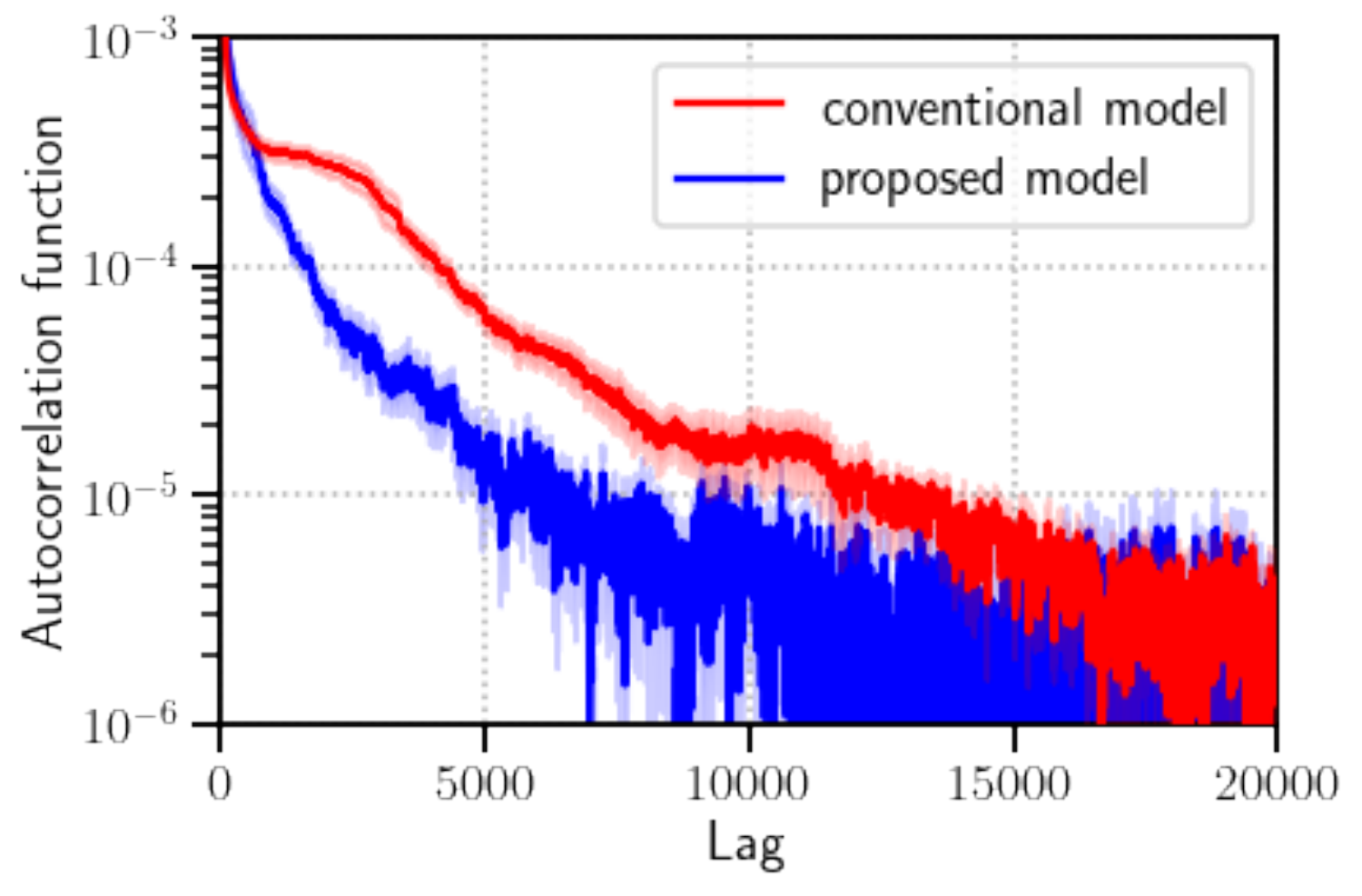}
   \label{3_b}
  }
  \caption{Transition and autocorrelation of error function $\mathcal{E}_N(\theta)$. (a) Error function at each MCS. The red and blue lines show $\mathcal{E}_N(\theta)$ of the conventional and proposed models, respectively. The shade represents the standard deviation for 10 independent runs. (b) The red and blue lines show the autocorrelation functions of the proposed and conventional models, respectively. The shade represents the standard error for 10 independent runs.} 
  \label{conv}
 \end{center}
\end{figure}

 \begin{figure}[t]
 \begin{center}
  \subfigure[]{
   \includegraphics[width=.47\columnwidth]{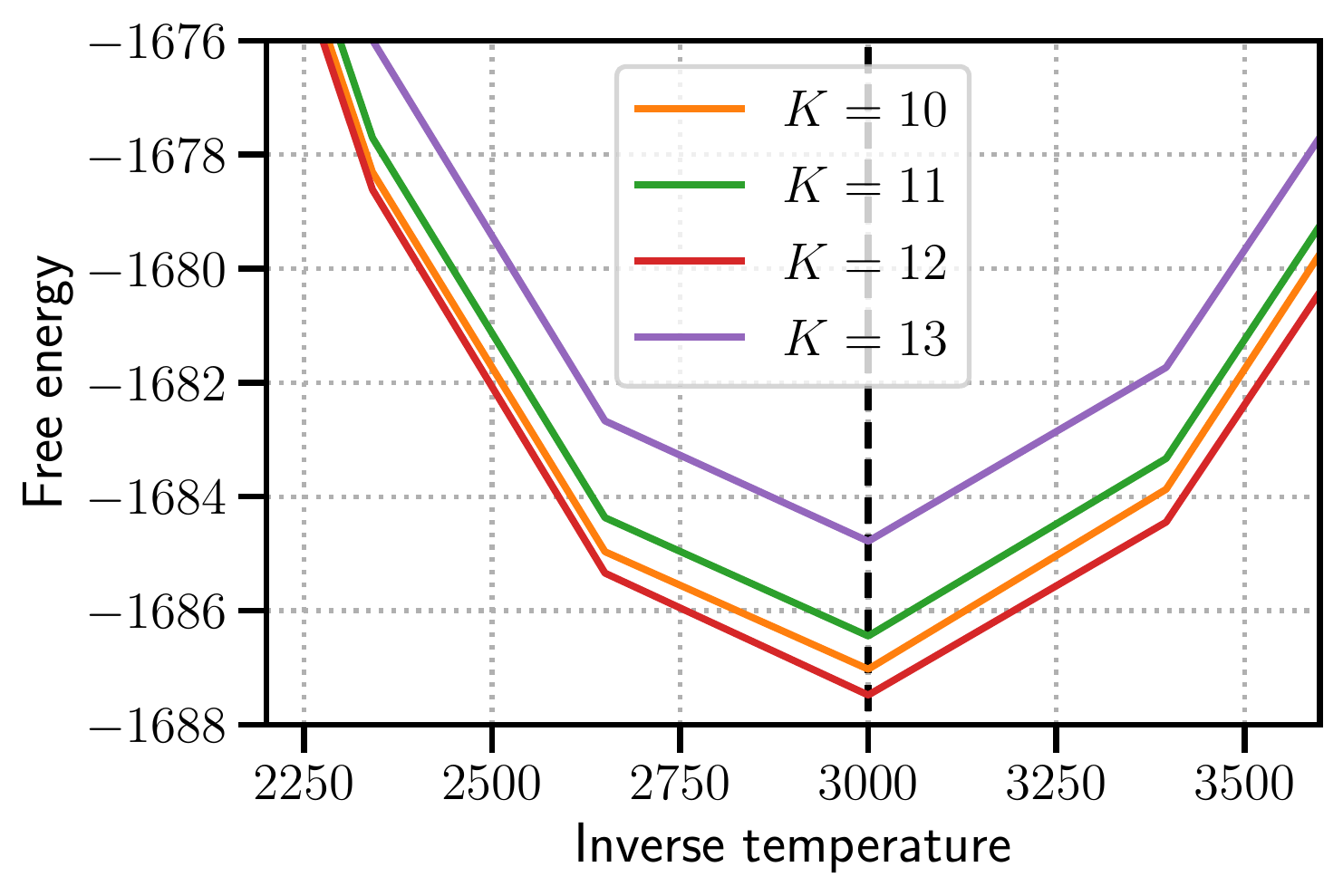}
  }~
  \subfigure[]{
   \includegraphics[width=.47\columnwidth]{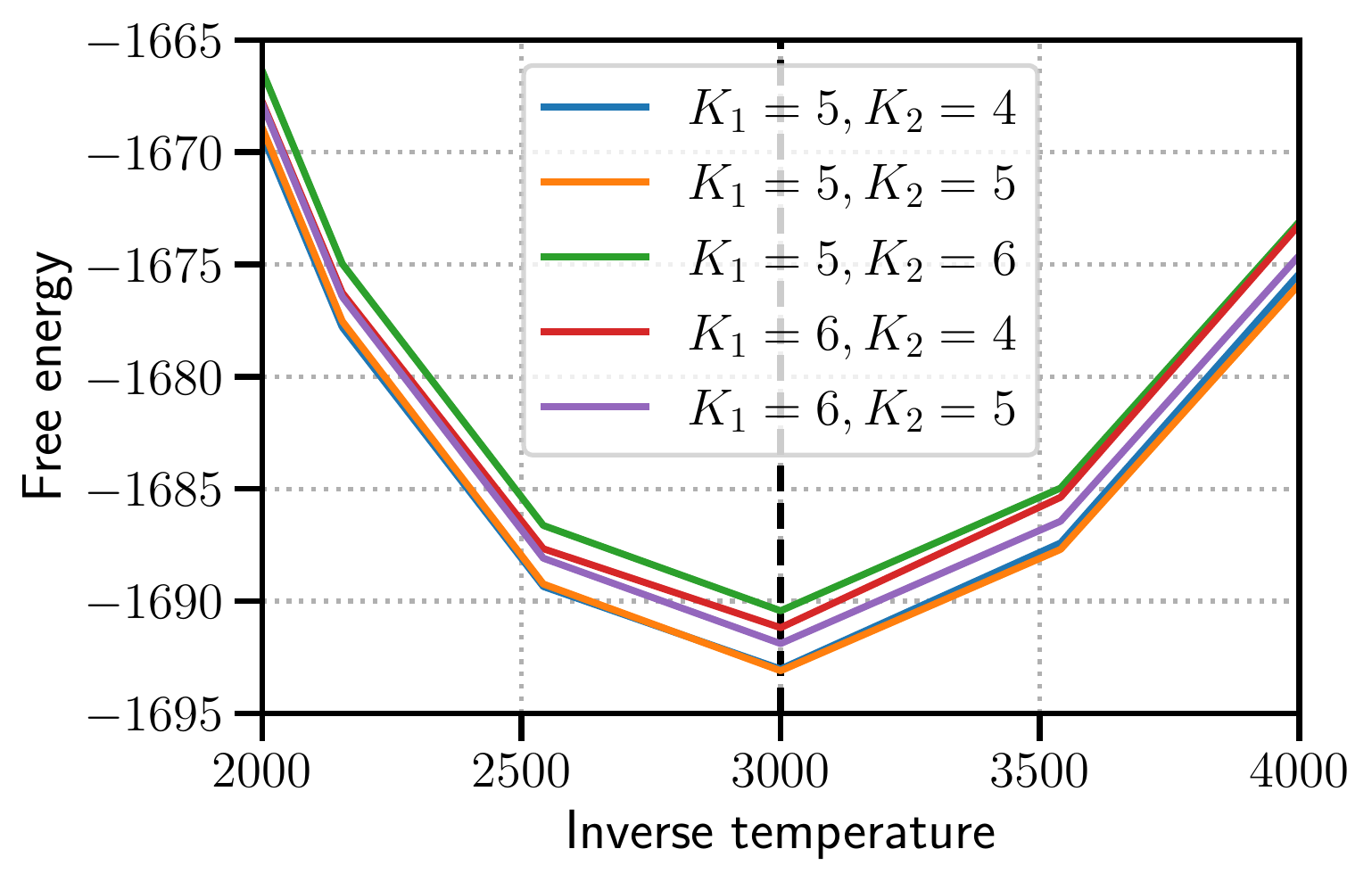}
  }
  \caption{Bayesian free energy as a function of inverse temperature $b$ of (a) conventional model $F_N(K,b)$ and (b) proposed model $F_N(K_1, K_2,b)$. The vertical black dashed lines show the true value $b = 3000$.} 
  \label{four}
 \end{center}
\end{figure}

\begin{figure}[t]
 \begin{center}
   \subfigure[]{
   \includegraphics[width=.47\columnwidth]{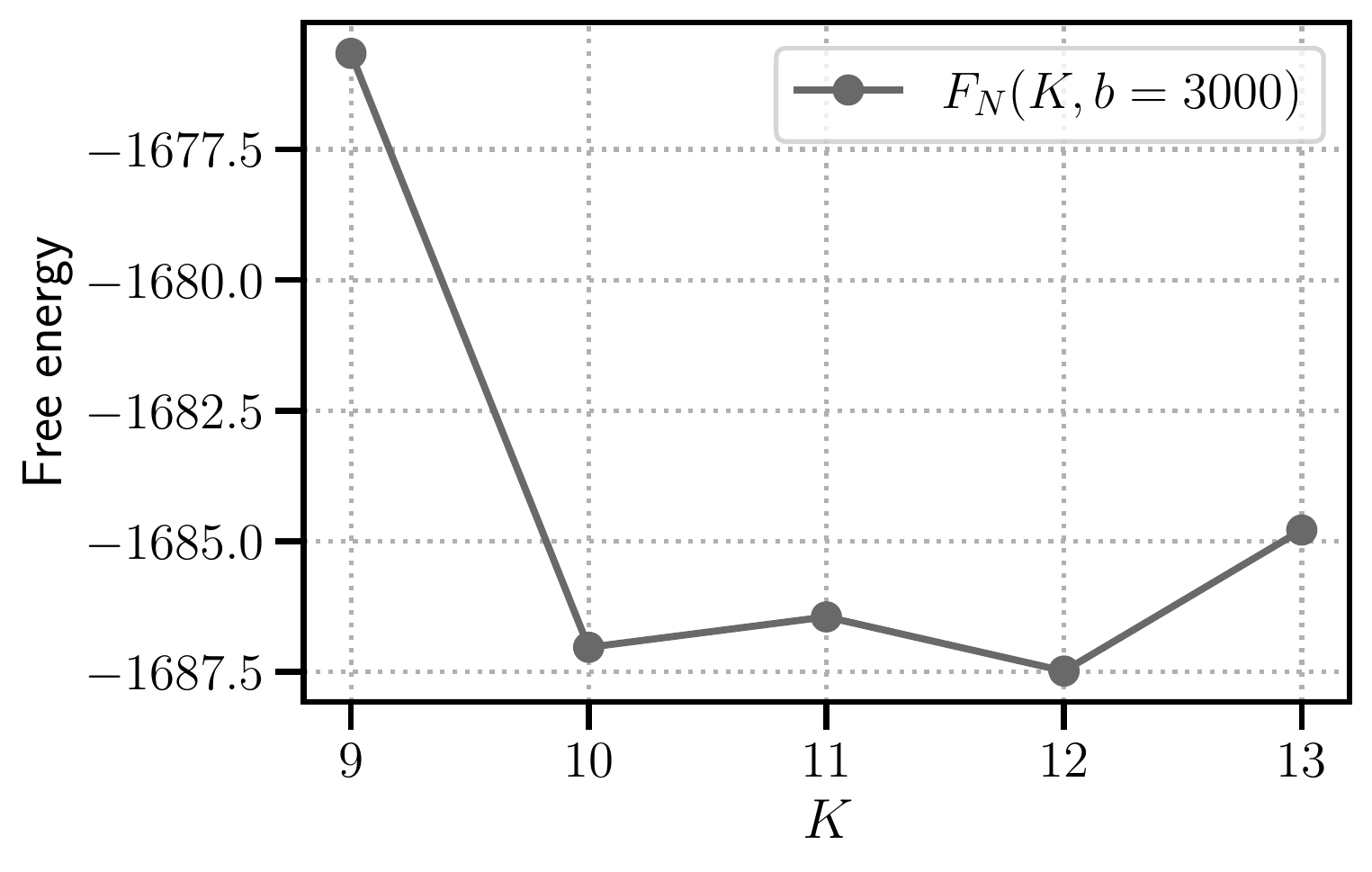}
   \label{5_a}
  }~
  \subfigure[]{
   \includegraphics[width=.47\columnwidth]{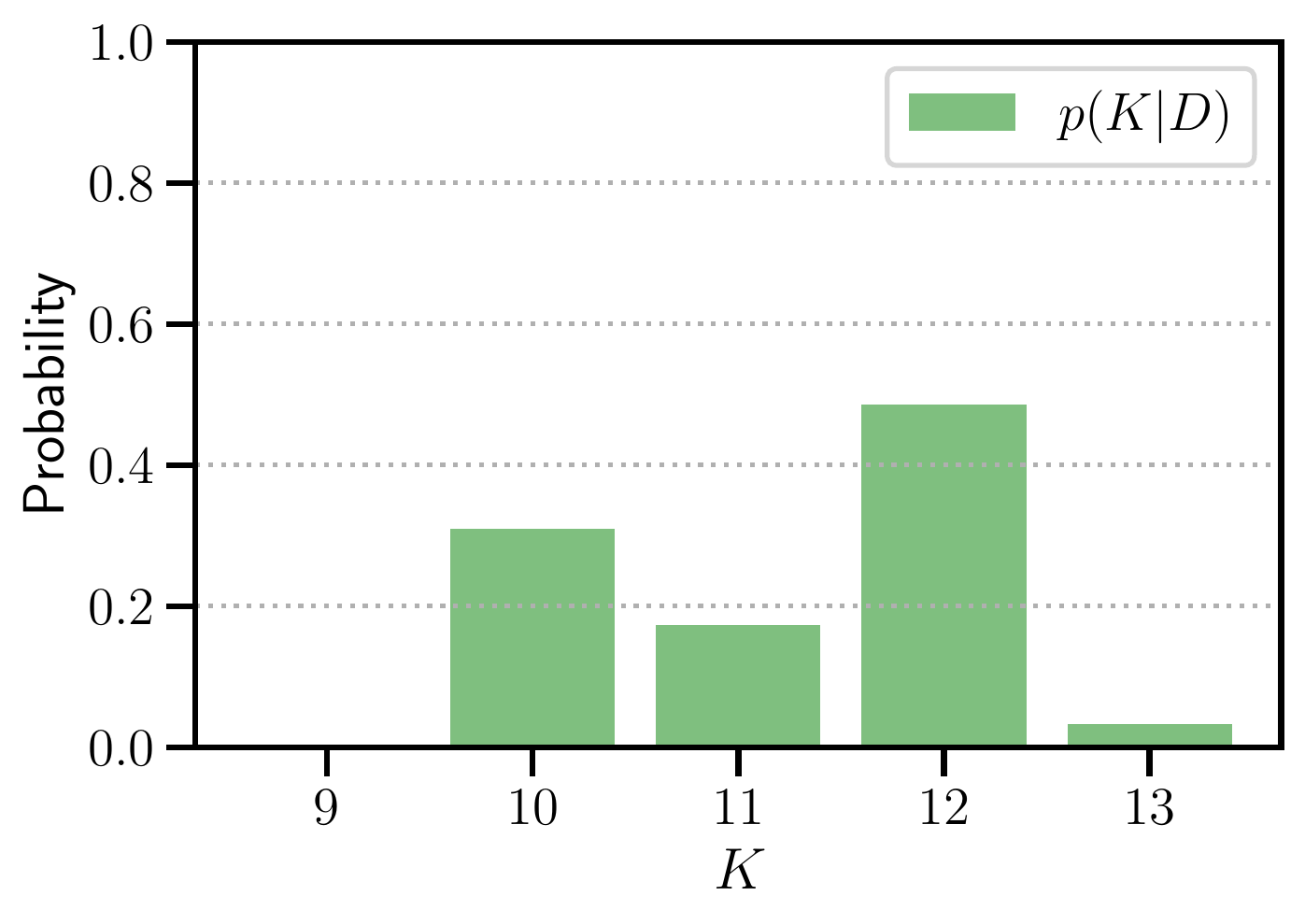}
   \label{5_b}
  }
  \subfigure[]{
   \includegraphics[width=.47\columnwidth]{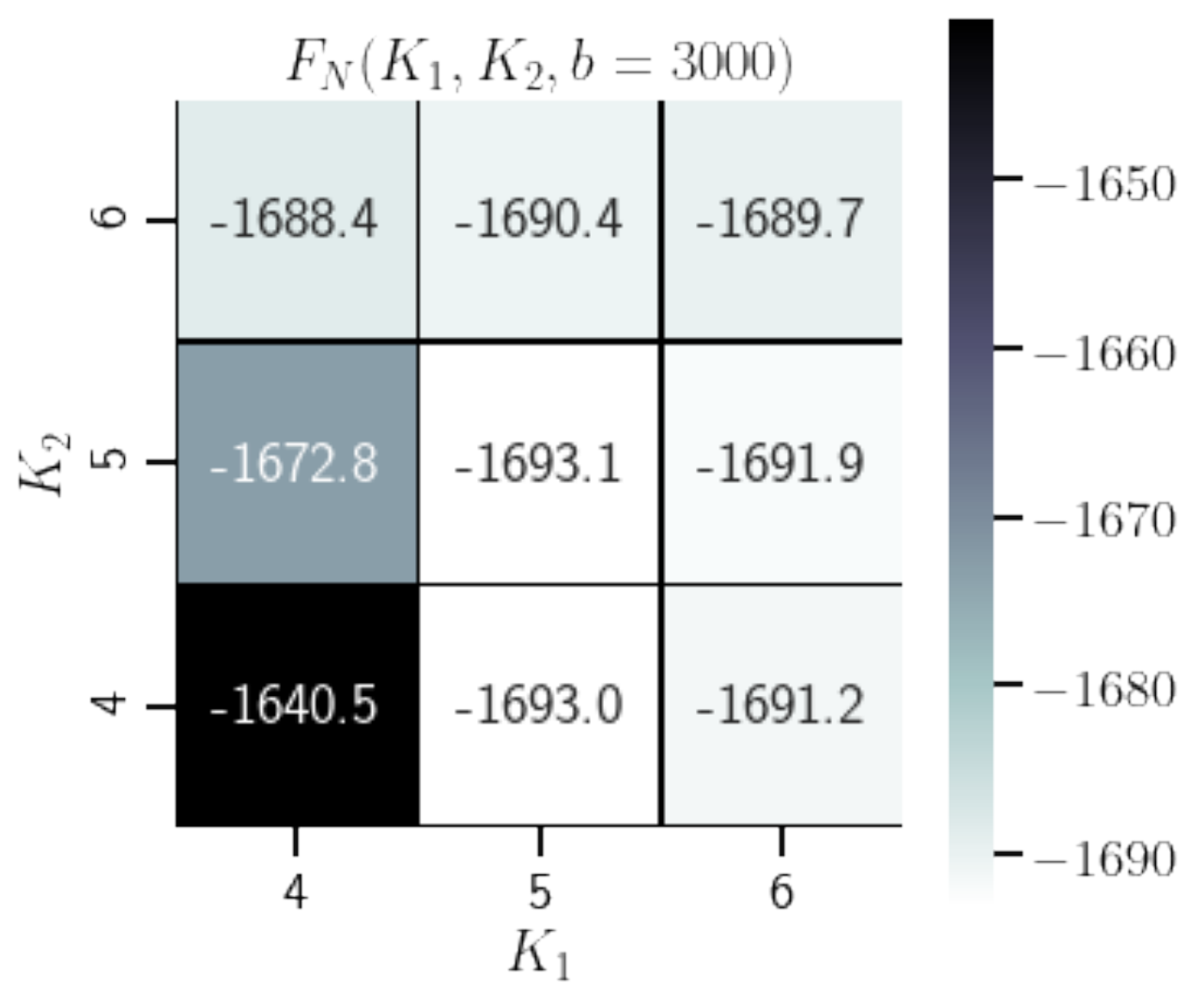}
   \label{5_c}
  }~
  \subfigure[]{
   \includegraphics[width=.47\columnwidth]{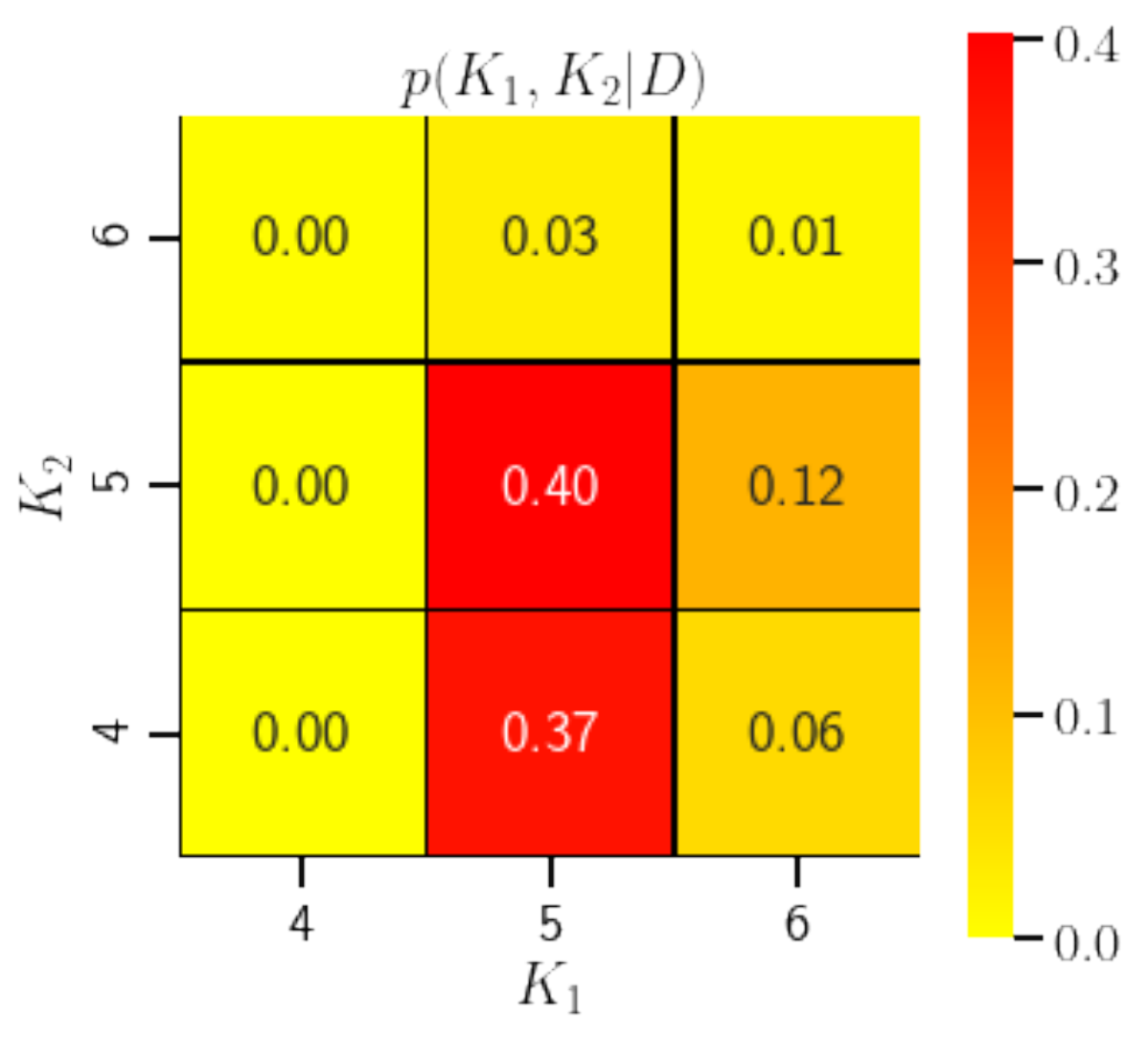}
   \label{5_d}
  }
  \caption{Bayesian free energy and probability of number of peaks for conventional and proposed models. (a) Free energy $F_N(K,b=3000)$. (b) Probability $p(K|D)$. (c) Free energy $F_N(K_1, K_2, b = 3000)$. (d) Probability $p(K_1, K_2  |D)$.} 
  \label{five}
 \end{center}
\end{figure}

First, we compare the proposed and conventional models in terms of convergence speed and sampling efficiency.
We set the number of peaks as $K=10$ for the conventional model and $(K_1, K_2) = (5,5)$ for the proposed model.
The number of replicas and inverse temperatures were set as $\xi = 1.18$ and $L = 92$ for both models, respectively.
 Figure \ref{3_a} shows the error function $\mathcal{E}_N(\theta)$ against MCSs. 
The initial value of each state $\theta_l$ for the EMC method was randomly chosen from the prior density $p(\theta|\mathcal{K})$. As seen from these results, the simulation using the proposed model can converge faster than that using the conventional model. Figure \ref{3_b} shows the autocorrelation function\cite{newman} of $\mathcal{E}_N(\theta)$. We can see that the simulation of the proposed model has a lower autocorrelation than that of the conventional model. From these results, we can say that the simulation using the proposed model can converge faster and obtain low-correlation samples in a small iteration compared with the conventional model. 

In the following, we show the results of the estimation of noise variance, number of peaks, and spectral parameters.
For inverse temperatures, we set $\xi = 1.18$ and $L = 92$ for the proposed model and $\xi = 1.132$ and $L = 120$ for the conventional model.
Figure \ref{four} shows the free energies $F_N(K,b)$ and $F_N(K_1, K_2, b)$ calculated as described in Sect. 2.4. As seen from these results, the minimum point of the free energy as a function of $b$ is  $\hat{b} = 3000$. According to these results, the noise estimation via Eq. (2.10) seems to obtain the correct noise variance.  
 \begin{figure}[t]
 \begin{center}
  \subfigure[]{
   \includegraphics[width=.47\columnwidth]{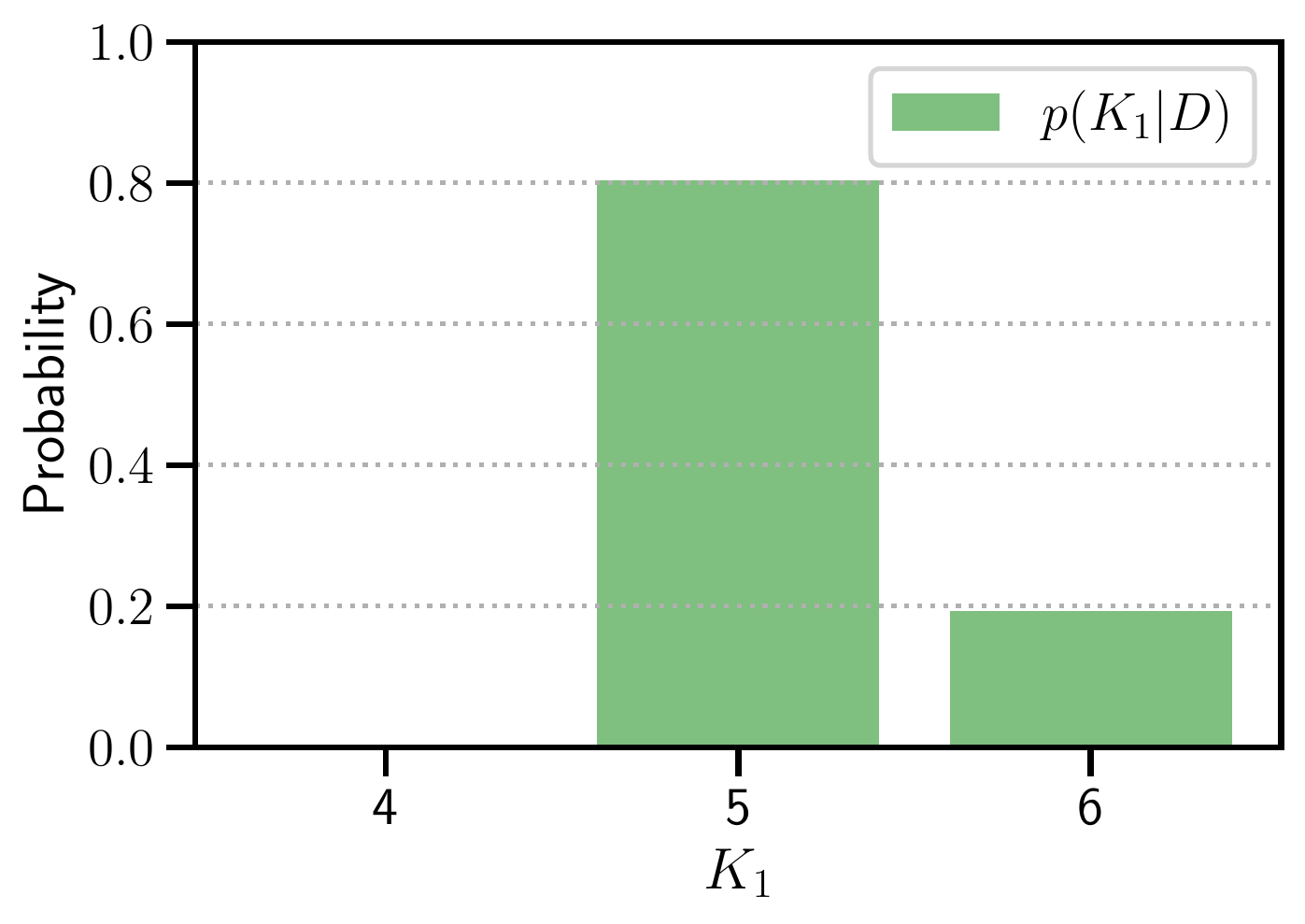}
  }~
    \subfigure[]{
   \includegraphics[width=.47\columnwidth]{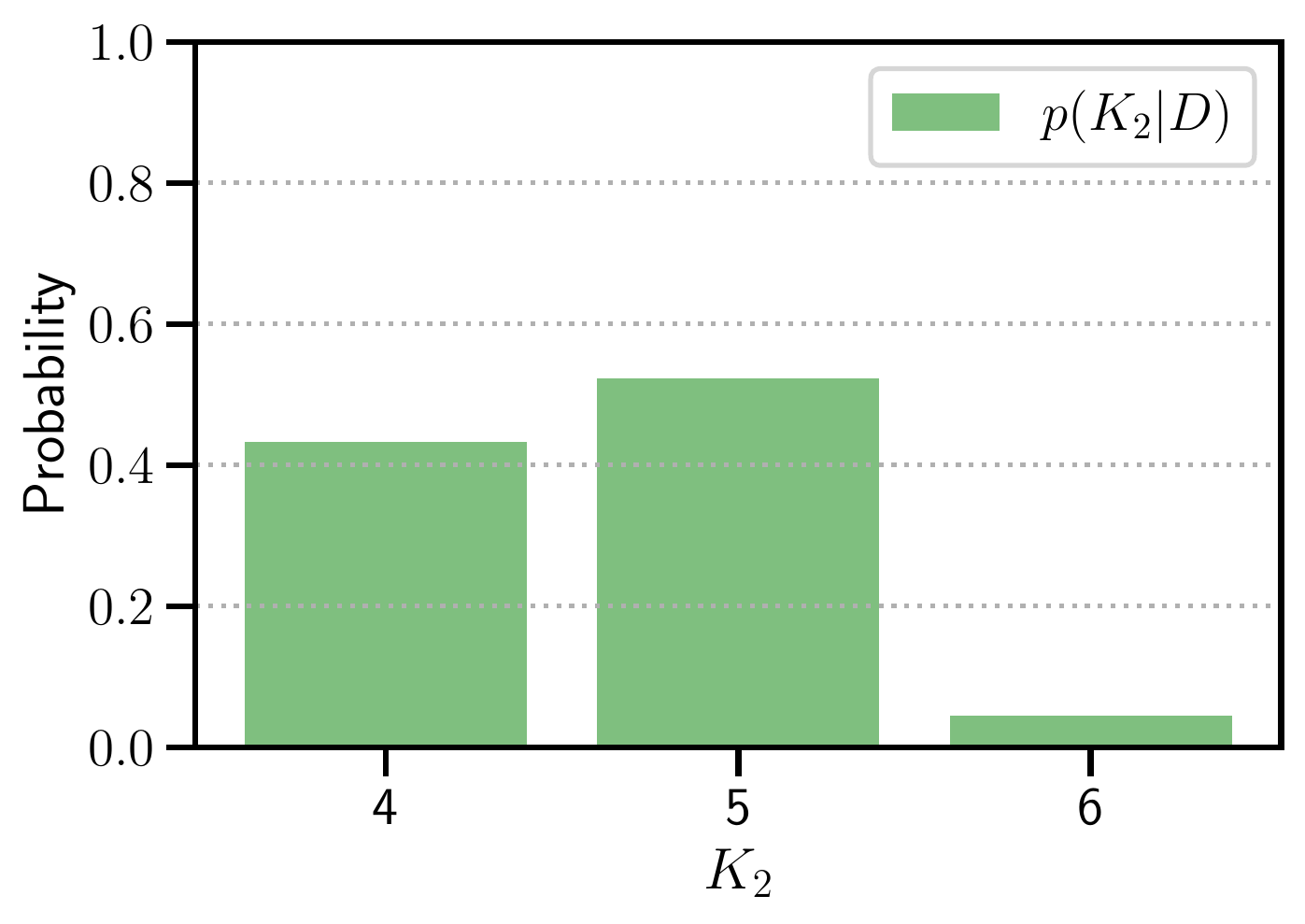}
  }
  \caption{Probability of the number of peaks below and above the absorption edge. (a)$p(K_1 | D)$ and (b) $p(K_2 | D)$.} 
  \label{six}
 \end{center}
\end{figure}
\begin{figure}[t]
 \begin{center}
  \subfigure[]{
   \includegraphics[width=120mm]{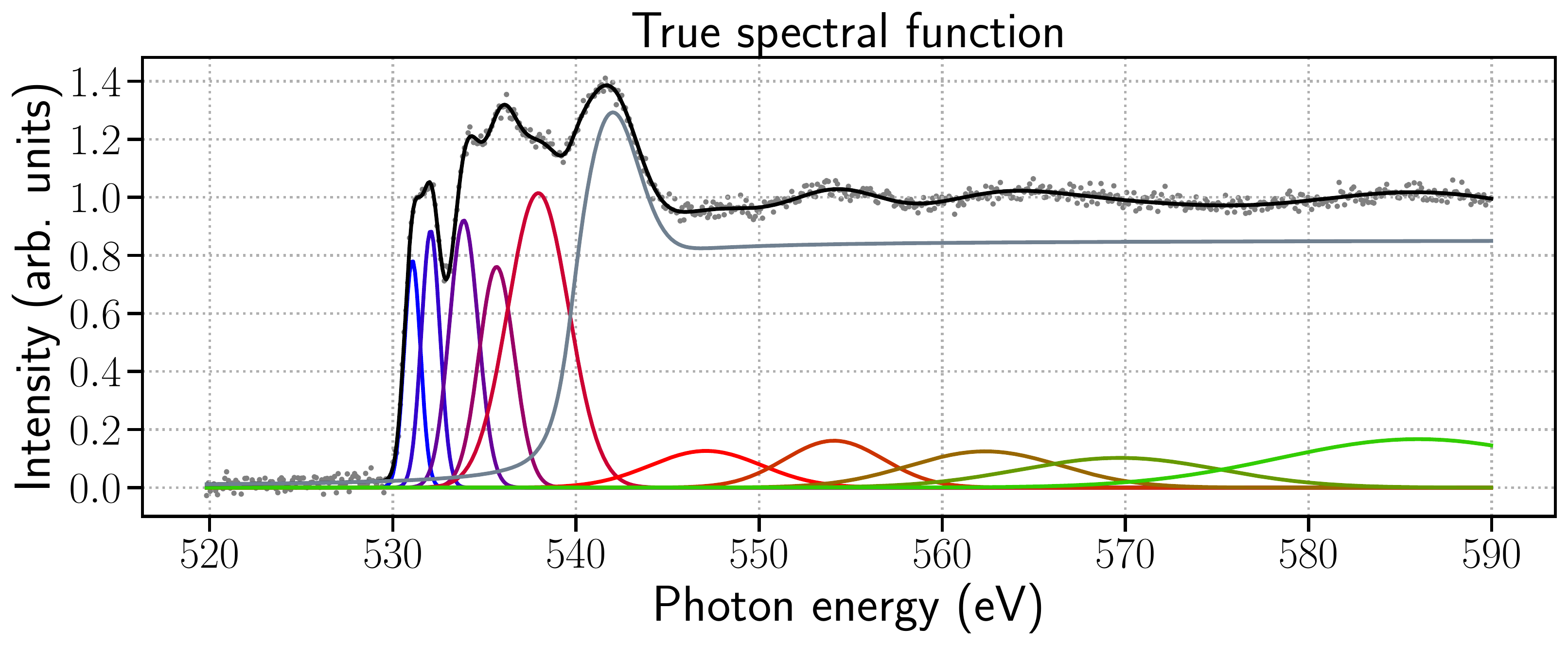}
   \label{7_a}
  }
  \subfigure[]{
   \includegraphics[width=120mm]{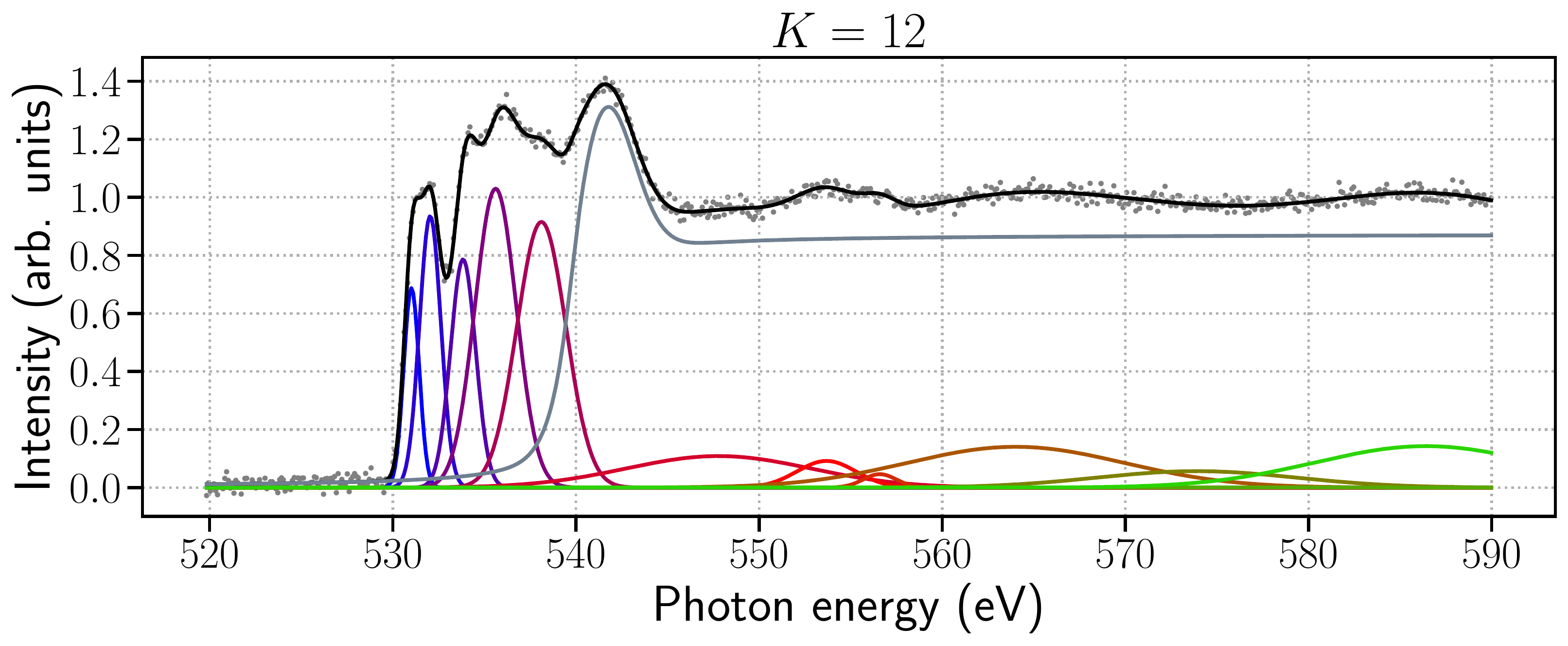}
   \label{7_b}
  }
    \subfigure[]{
   \includegraphics[width=120mm]{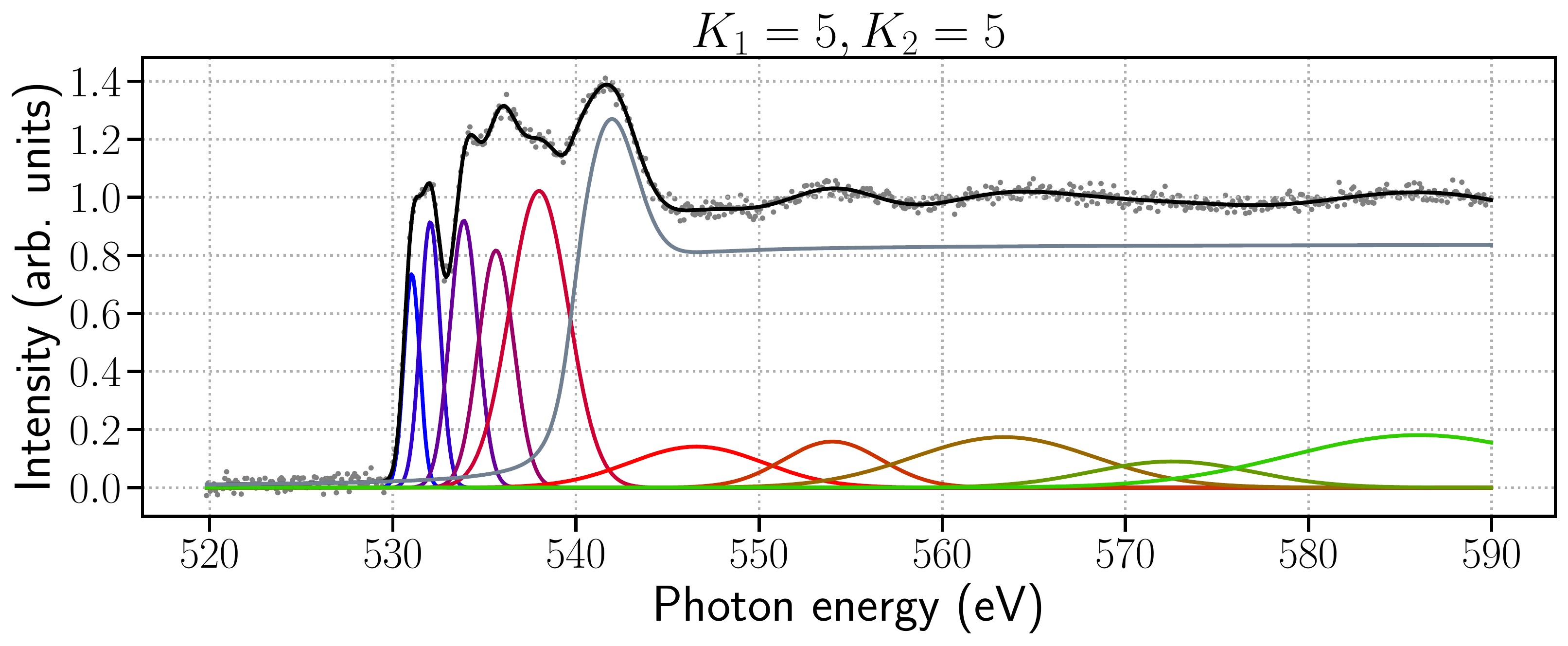}
   \label{7_c}
  }
  \caption{True synthetic spectra and results of MAP fitting. (a) True spectral model and synthetic data. (b) Fitting result of conventional model for the estimated number of $K=12$ peaks. (c) Fitting result of proposed model for the estimated number of $(K_1, K_2) = (5,5)$ peaks. }  
  \label{seven}
 \end{center}
\end{figure}

Figure \ref{five} shows the results of the model selection. Figures \ref{5_a} and \ref{5_b} show the results of $F_N(K, b = 3000)$ and $p(K|D)$, respectively. From these results, the estimation by the conventional method yields an incorrect number $K$ of peaks, $K=12$. Figures \ref{5_c} and \ref{5_d} show the results of $F_N(K_1, K_2, b = 3000)$ and $p(K_1, K_2|D)$ respectively. From these results, the estimation by the proposed method yields an appropriate number $(K_1, K_2)$ of peaks, $(K_1, K_2) = (5,5)$. As seen from Fig. \ref{five}, both the free energy $F_N(K_1, K_2, b = 3000)$ and the probability density $p(K_1, K_2|D)$ have similar values between the model of $(K_1, K_2) = (5,5)$ and $(K_1, K_2) = (5,4)$. 

For a detailed discussion, we show the probability densities $p(K_1 | D)$ and $p(K_2 | D)$ in Fig. \ref{six}. Although the probability $p(K_1  =5|D)$ has a significantly large value, the difference between $p(K_2 = 5 | D)$ and $p(K_2 = 4 | D)$ is small. 
Since the low-energy domain is the primary consideration in correspondence with first-principles calculations, it is an advantage of the proposed model to obtain more significant results for $K_1$.

Figure \ref{seven} shows the results of the MAP fitting with the estimated number of peaks, $K=12$ and $(K_1, K_2) = (5,5)$. As seen from Fig. \ref{7_b}, the result of the conventional model is inappropriate although the peak positions in the low-energy region seem to be estimated correctly.
However, the proposed model can estimate the overall structure of the true model correctly, as seen from Fig. \ref{7_c}.
Thus, we can estimate the values of the number of peaks and the position, intensity, and width of each peak by using the proposed model.

In addition, we see that the values of free energy of the proposed model are lower than those of the conventional model. In other words, the proposed model has higher model evidence than the conventional model.

\section{Discussion and Conclusion}
In this paper, we proposed a model that discriminates between the low- and high-energy domains.
Through numerical experiments, we obtained the following observations.
First, the proposed model is superior to the conventional one in terms of convergence speed and sampling efficiency.
Second, the proposed model showed better accuracy than the conventional one in the model selection and the MAP estimation.
Third, the proposed model has a lower free energy than the conventional one.
Fourth, the proposed model enables us to estimate the number of peaks focusing on the important energy domain in terms of material science.

In this paper, we validated our framework via synthetic data analysis.
Future work will consist of real data analysis and comparison of the analysis results with first-principles calculations.
Our method will be used for this as well and will contribute to materials science.

Futhermore, superior performances of the proposed model result from the design of prior distributions. We also showed the importance of using physical knowledge to design prior distributions.
Our framework is applicable to not only the analyses of XANES but those of any other spectra with energy-dependent structures.

\begin{acknowledgments}
The authors would like to express their sincere gratitude to Koki Okajima for his insightful comment.
Their sincere thanks also go to Koki Obinata for valuable advice and discussion.
This work was supported by JST CREST (Grant Nos. JPMJCR1761 and JPMJCR1861) from the Japan Science and Technology Agency (JST).
\end{acknowledgments}

\end{document}